\documentclass[12pt]{article}

\pdfoutput=1
\usepackage{jheppub}

\newcommand{\C}{{\mathbb C}}

\DeclareMathOperator{\Bun}{Bun}

\newcommand{\no}[1]{:\mathrel{#1}:}

\title{Vertex Algebra constructions for (analytic) Geometric Langlands in genus zero}
\author[1]{Davide Gaiotto}
\affiliation[1]{Perimeter Institute for Theoretical Physics, Waterloo, Ontario, Canada N2L 2Y5}

\begin{document}
\abstract{We employ two-dimensional chiral algebra techniques to produce solutions of certain differential and integral equations
which occur in the context of the Analytic Geometric Langlands Program. In particular, we build ``Multiplication Kernels'' $K_3(x,x',x'')$ which intertwine 
the action of $\mathfrak{sl}_2$ Gaudin Hamiltonians on three sets of variables.}
\maketitle
\flushbottom
\section{Introduction}
The objective of this paper is to employ the 2d chiral algebra of {\it symplectic bosons} to produce solutions of certain differential and integral equations
which occur in the context of the Analytic Geometric Langlands Program \cite{Teschner:2017djr,Etingof:2019pni,Etingof:2021eub,Etingof:2021eeu,Gaiotto:2021tsq}.
The general strategy we employ was delineated in \cite{Gaiotto:2021tsq}, building on a construction from \cite{Gaiotto:2016wcv}. The main contribution of this paper is to provide the tools 
necessary to make contact with the ramified version of the Langlands program. This allows for non-trivial calculations in genus zero, where the differential and integral equations and their solutions can be described very explicitly. 

The genus zero setup is simple enough that we can state our basic results without introducing a complicated mathematical or physical background.
The differential equations involve the $\mathfrak{sl}_2$ Gaudin Hamiltonians
\begin{equation}
H_i = \sum_{j =0\,| j \neq i}^n \frac{e_i f_j + f_i e_j + 2 h_i h_j}{z_i-z_j}
\end{equation}
built from a collection of differential operators 
\begin{equation} \label{eq:basic}
f_i = \partial_{x_i} \qquad \qquad h_i = x_i \partial_{x_i} + \frac12 \qquad \qquad e_i = - x_i^2 \partial_{x_i} - x_i 
\end{equation}
which satisfy the commutation relations of $\mathfrak{sl_2}$ for all $i=1, \cdots n$. Here $n$ is an integer greater than $2$ and the $z_*$ variables represent some fixed collection of $n$ distinct complex parameters. We consider three copies $H_i$, $H'_i$, $H''_i$ of the Gaudin Hamiltonians, acting on three different sets of variables $x_*$, $x'_*$, $x''_*$ with the same values of the $z_*$'s. 

Our objective is to produce a solution $K_3(x_*,x_*',x_*'';z_*)$ of the intertwining relations
\begin{equation}\label{eq:intertwine}
\boxed{H_i \circ K_3 = H'_i \circ K_3 = H''_i \circ K_3}
\end{equation}
which is also invariant under the simultaneous $\mathfrak{sl_2}$ rotation of all the $x_*$:
\begin{equation}\label{eq:rotate}
\sum_{i=0}^n e_i \circ K_3 = 0 \qquad \qquad \sum_{i=0}^n h_i \circ K_3 = 0 \qquad \qquad \sum_{i=0}^n f_i \circ K_3 = 0
\end{equation}
as well as simultaneous $\mathfrak{sl_2}$ rotation of all the $x'_*$ or all the $x''_*$.

The (multivalued) solution we produce is 
\begin{equation}
K_3(x_*,x_*',x_*'';z_*) = \frac{1}{\sqrt{\det A}}
\end{equation}
where the $n \times n$ matrix $A$ has entries
\begin{align}
A_{ij} &= \frac{(x_i-x_j)(x'_i-x'_j)(x''_i-x''_j)}{z_j-z_i} \qquad \qquad i \neq j \cr
A_{ii} &=0 \, .
\end{align}
We will refer to this solution as the ``chiral kernel''.

The relations (\ref{eq:rotate}) are the infinitesimal version of a covariance property 
\begin{equation}
K_3 \to K_3 \prod_i (c x_i + d) 
\end{equation}
under fractional linear transformations 
\begin{equation}
x_i \to \frac{a x_i + b}{c x_i + d} \,.
\end{equation}
It follows immediately from the transformation rule
\begin{equation}
A_{ij} \to \frac{A_{ij}}{(c x_i + d)(c x_j + d)} \, .
\end{equation}

Our proof of the relations (\ref{eq:intertwine}) will employ a realization of $K_3$ as a sphere correlation function of $n$ vertex operators in an appropriate 2d chiral algebra
\begin{equation}
V_{\mathfrak{sl}_2,3} \equiv \mathrm{Sb}[\mathbb{C}^2 \otimes\mathbb{C}^2 \otimes\mathbb{C}^2]
\end{equation}
consisting of eight free symplectic boson, aka four $\beta \gamma$ systems of scaling dimension $\frac12$. 
This chiral algebra includes three copies of the $\mathfrak{sl}(2)_{-2}$ Kac-Moody algebra at critical level. In turn, each copy contains a bilinear Sugawara element $S(z)$ which is central, i.e. has trivial OPE with all the currents. 

By construction, the insertion of the Sugawara element (or any other operator built from the currents) 
in any correlation function can be expanded out with the help of the Kac-Moody Ward identities. The details of the expansion depend on the geometry of the problem and on the properties of the vertex operators involved. Schematically, we could write such a relation as 
\begin{equation}
\langle S(z) \cdots \rangle = H_S(z) \circ \langle \cdots \rangle
\end{equation}
with the operator $H_S(z)$ representing the solution of the Ward identities. It is easy to argue that $[H_S(z), H_S(z')]=0$. In appropriate contexts, $H_S(z)$ becomes a generating function of Gaudin or Hitchin's Hamiltonians.

A special property of $V_{\mathfrak{sl}_2,3}$ is that the Sugawara elements for the three sets of currents coincide: $S(z) = S'(z) = S''(z)$. 
This implies that any correlation functions satisfy intertwining relations of the form 
\begin{equation}
\boxed{H_S(z) \circ \langle \cdots \rangle = H_{S'}(z) \circ \langle \cdots \rangle = H_{S''}(z) \circ \langle \cdots \rangle}
\end{equation}

The main ingredient of our construction is the definition of certain vertex operators ${\cal V}(x,x',x'';z)$ for $V_{\mathfrak{sl}_2,3}$ 
which are highest weight for each of the three Kac-Moody algebras, with the zeromodes of the currents acting as the 
$e$, $h$ and $f$ vectorfields on the variables $x$, $x'$ and $x''$ respectively. 

The Ward identities for a sphere correlation function of $n$ highest weight vertex operators express a Sugawara vector insertion 
as a linear combination of the Gaudin Hamiltonians: 
\begin{equation}
\left\langle  S(z) \prod_{i=1}^n {\cal V}(x_i,x_i',x_i'';z_i) \right\rangle= \left[\sum_i \frac{1}{z-z_i} H_i\right] \circ \left\langle \prod_{i=1}^n {\cal V}(x_i,x_i',x_i'';z_i) \right\rangle
\end{equation}

The chiral kernel is produced by an explicit evaluation of the sphere correlation function:
\begin{equation}
K_3(x_*,x_*',x_*'';z_*) \equiv \left\langle \prod_{i=1}^n {\cal V}(x_i,x_i',x_i'';z_i) \right\rangle_{\mathbb{CP}^1}
\end{equation}
and thus satisfies the relations (\ref{eq:intertwine}). 

\subsection{The non-chiral kernels}
The 2d chiral algebra $V_{\mathfrak{sl}_2,3}$ is not quite a fully fledged 2d CFT. It can be at best understood as a {\it relative} 2d theory and produces correlation functions which are not single-valued functions.\footnote{The partition function or correlation functions of a relative theory are not numbers, or functions on the space of parameters. Instead, they are elements of a vector space which depends on the topology of spacetime and is often fibered non-trivially on the space of parameters. The term ``conformal block'' is sometimes employed to denote these relative correlation functions, but the term is overloaded. In this paper we will simply use the term ``correlation function'' both for relative and standard theories.} This problem is already visible in the sign ambiguity of $K_3$ and becomes more serious for other correlation functions we will consider below. 

Elaborating on \cite{Gaiotto:2021tsq}, we will introduce a fully-fledged (i.e. not relative) 2d CFT $T_{\mathfrak{sl}_2,3}$ whose correlation functions are single-valued combinations of correlation functions of $V_{\mathfrak{sl}_2,3}$ and of its anti-chiral analogue.
Correlation functions of $T_{\mathfrak{sl}_2,3}$ satisfy Ward identities for 
both chiral and anti-chiral Kac-Moody currents and thus satisfy both holomorphic and anti-holomorphic versions of the intertwining relations. 

We define a non-chiral version $\mathbb{V}(x_i,x_i',x_i'';z_i)$ of the ${\cal V}(x_i,x_i',x_i'';z_i)$ vertex operators and compute a sphere correlation function. This gives a ``complex'' kernel
\begin{equation}
\mathbb{K}_3 \equiv \left\langle \prod_{i=1}^n \mathbb{V}(x_i,x_i',x_i'';z_i) \right\rangle_{\mathbb{CP}^1} = \frac{1}{|\det A|}
\end{equation}
which intertwines both the holomorphic and anti-holomorphic versions of the Gaudin Hamiltonians and is single-valued. 

The non-chiral theory $T_{\mathfrak{sl}_2,3}$ admits some simple boundary conditions and can also be defined on non-orientable surfaces. Correlation functions on a disk or a cross-cap satisfy Ward identities and intertwining relations which are a ``real'' analogue of these for a sphere. They may involve either bulk operators $\mathbb{V}(x_i,x_i',x_i'';z_i)$ or operators $\mathbb{V}^\partial(x_i,x_i',x_i'';z_i)$ defined only along the boundary. 

These correlation functions can be usually computed by a reflection trick, expressing them in terms of the chiral kernel. Each bulk vertex operator is mapped to a pair of chiral vertex operators, 
at points related by a complex involution which may or not have fixed points depending on the surface being a disk or cross-cap, with parameters which depend on the selected boundary conditions.
Boundary vertex operators give a single chiral vertex operator at ``real'' positions fixed by the involution. The resulting ``real'' kernels take the general form 
\begin{equation}
\mathbb{K}^{\mathbb{R}}_3 = \frac{\theta(\det A)}{\sqrt{\det A}}
\end{equation}
With $\theta$ being the Heaviside step function. Here the parameters of the chiral operators produced by the reflection trick guarantee that $A$ satisfies certain reality conditions which imply that $\det A$ is real. 

\subsection{Hecke operators intertwining relations}
The theory $T_{\mathfrak{sl}_2,3}$ has a large symmetry group consisting of locally holomorphic, position-dependent $PGL(2) \times PGL(2) \times PGL(2)$ 
transformations. The most familiar consequence of this symmetry is the existence of Kac-Moody currents and their associated Ward identities. These encode the infinitesimal symmetries of the system. 

A less well-known fact is that invariance under these symmetry transformations leads to a larger set of ``spectral flow'' Ward identities. These control the 
insertion of vertex operators which do not belong to the usual Kac-Moody vacuum module but 
are created by singular gauge transformations. From that perspective, the Hecke operators ${\cal H}_{\cal S}(z)$ which appear in the analytic GL program simply encode the solutions 
\begin{equation}
\langle {\cal S}(z) \cdots \rangle = {\cal H}_{\cal S}(z) \circ \langle \cdots \rangle
\end{equation}
of spectral flow Ward identities for some particularly nice central vertex operators ${\cal S}(z)$. The Hecke operators ${\cal H}_{\cal S}(z)$ take the form of integral operators simply because 
${\cal S}(z)$ is constructed as an integral transform of simpler vertex operators created by specific singular gauge transformations. \footnote{Similar relations can be written in the chiral theory $V_{\mathfrak{sl}_2,3}$ as well, but are complicated by the multi-valuedness on the two sides. The operator on the right hand side involves a contour integral and different choices of integration contours give different versions of the left hand side. These structures are understood from the standard categorical Geometric Langlands program.}

As discussed in \cite{Gaiotto:2021tsq} and expanded further here, $T_{\mathfrak{sl}_2,3}$ has the special property that the vertex operators associated to Hecke operators for the three $SL(2)$ groups actually coincide: ${\cal S}(z) = {\cal S}'(z) = {\cal S}''(z)$. 
As a consequence, the non-chiral kernels intertwine the action of three sets of Hecke operators:
\begin{equation}
\boxed{{\cal H}_{\cal S}(z) \circ \langle \cdots \rangle = {\cal H}_{{\cal S}'}(z) \circ \langle \cdots \rangle = {\cal H}_{{\cal S}''}(z) \circ \langle \cdots \rangle}
\end{equation}

\subsection{General weights}
The construction we have just sketched can be extended naturally to allow for non-trivial weights in the $\mathfrak{sl}_2$ generators:
\begin{equation} \label{eq:basic}
f_i = \partial_{x_i} \qquad \qquad h_i = x_i \partial_{x_i} -j_i \qquad \qquad e_i = - x_i^2 \partial_{x_i} + 2 j_i x_i 
\end{equation}
The corresponding chiral kernel is given by an integral formula of the schematic form
\begin{equation} \label{eq:wei}
K_3(x_*,x_*',x_*'';z_*) = \oint \prod_i \left[ \lambda_i^{-2j_i-1} d \lambda_i\right] e^{-\lambda^t A \lambda}
\end{equation}

Notice that the contour integral admits now a variety of different contours. Single-valued non-chiral kernels $\mathbb{K}_3$ or $\mathbb{K}^{\mathbb{R}}_3$ can be defined respectively by 
a complex integral 
\begin{equation}
\mathbb{K}_3(x_*,x_*',x_*'';z_*) = \int \prod_i \left[ \lambda_i^{-2j_i-1} d \lambda_i\right]\left[ \bar \lambda_i^{-2 \bar j_i-1} d \bar \lambda_i\right] e^{-\lambda^t A \lambda+ \mathrm{c.c.}}
\end{equation}
or by a real choice of contour in equation (\ref{eq:wei}). In either cases this requires certain constraints on the weights $(j_i, \bar j_i)$ or $j_i$. 

\subsection{Generalizations}
Consider some 2d chiral algebra $A$ which includes Kac-Moody currents at critical level for some reductive group $G$. The inclusion gives a map from the whole Kac-Moody chiral algebra to $A$, mapping the center of the Kac-Moody algebra to central elements in $A$. The map from the center of the Kac-Moody algebra may have a kernel. Applying the Kac-Moody Ward identities to elements in the kernel we discover that the partition function or correlation functions for $A$ will be annihilated by the $G$ quantum Hitchin Hamiltonians which express the solution to the Ward identities. 

If we can build non-relative 2d CFTs out of $A$, the partition function or correlation functions of these CFTs will give single-valued kernels which are annihilated by these quantum Hitchin Hamiltonians and thus may have applications to the analytic geometric Langlands program. If the correlation functions are sufficiently algebraic, the kernels may even have applications outside the geometric setting \cite{2007math......2206K}. 

Gauge theory constructions provide many examples of such interesting chiral algebras and 2d CFTs. The corresponding kernels may be more or less computable, 
depending on the complexity of the chiral algebras. Weakly-coupled gauge theory constructions produce either free chiral algebras or BRST reductions of free chiral algebras. The former give kernels which can be readily computed. The latter require more work, but should still be useful. 

Unfortunately, many interesting gauge-theory constructions involve strongly-coupled theories. This includes the the analogue of $V_{\mathfrak{sl}_2,3}$
for other Lie algebras, giving kernels which intertwine three copies of a full set of quantum Hitchin Hamiltonians and Hecke operators. Perhaps surprisingly, the vertex algebras themselves can be built explicitly \cite{2018arXiv181101577A}, but the full 2d CFTs are not yet understood and even the calculation of correlation functions is challenging. 

The simplest free example is a collection of $4N(N+1)$ symplectic bosons, treated as a bifundamental representation of $Sp(2N) \times SO(2N+2)$. 
This impose a constraint on the Hitchin Hamiltonians which is Langlands dual to the embedding $SO(2N+1) \subset SO(2N+2)$. We will look at that example in some detail. We will also discuss in some detail the chiral algebra $V_{\mathfrak{sl}_2,4}$ which produces kernels intertwining 
four copies of a full set of quantum Hitchin Hamiltonians and Hecke operators for $PSU(2)$.

\subsection{Structure of the paper}
In Section \ref{sec:symp} we give a general discussion of the theory of symplectic bosons. We review its basic properties and introduce some unusual vertex operators both in the Neveu-Schwarz and Ramond sectors. We describe the calculation of genus zero correlation functions in the presence of these vertex operators. We discuss the effect of singular gauge transformations. We briefly discuss some boundary conditions. In Section \ref{sec:trifu} we specialize the previous discussion to $V_{\mathfrak{sl}_2,3}$. We map the parameters of vertex operators to the space of parabolic bundles and describe the chiral algebra interpretation of the Gaudin Hamiltonians. We introduce a special Ramond vertex operator which represents the action of a minimal Hecke operator. We present some sample calculations for the four-punctured sphere. 
In Section \ref{sec:emb} we briefly discuss an interesting kernel associated to $\mathfrak{sp}(2N) \times \mathfrak{so}(2N+2)$ Gaudin Hamiltonians. In Section \ref{sec:brst} we discuss the non-chiral version of the BRST reduction and its application to the construction of multiplication kernels for four or more sets of variables. 

\section{Symplectic bosons} \label{sec:symp}
Although some of our calculations are very specific to $V_{\mathfrak{sl}_2,3}$ and its close relatives, some aspects apply to 
symplectic bosons in general and may be novel. This section is thus devoted to a general review of chiral and non-chiral symplectic bosons.

\subsection{Action, currents, genus $0$ correlation functions, non-chiral theory.}
Chiral symplectic bosons are the Grassmann-even analogue of 2d free chiral fermions: a $\beta \gamma$ systems where the 
$\beta$ and $\gamma$ fields both have scaling dimension $\frac12$. This particular choice of dimensions puts the two fields on an even footing 
and endows the theory with an $SL(2)$ global symmetry rotating them into each other. Similarly, if we take $n$ copies of the elementary $\beta \gamma$ systems
with scaling dimensions $\frac12$ we obtain a theory with $Sp(2n)$ global symmetry. 

We will collect all the fields into a $2n$-component vector $Z^a$ and write the action in an manifestly $Sp(2n)$-invariant form
\begin{equation} \label{eq:acbare}
\frac{1}{2 \pi} \int d^2 z\,\omega_{ab} Z^a \bar\partial Z^b
\end{equation}
where $\omega_{ab}$ is a symplectic form. 

The OPE can also be written in a manifestly $Sp(2n)$-invariant form
\begin{equation} \label{eq:ope}
Z^a(z) Z^b(w) \sim \frac{\omega^{ab}}{z-w}
\end{equation}
where $\omega^{ab}$ is the inverse symplectic form. 

The normal ordered bilinears
\begin{equation}
J^{ab}(z) \equiv \;\no{Z^a(z) Z^b(z)}
\end{equation}
define $\mathfrak{sp}(2n)_{-\frac12}$ Kac-Moody currents of level $-\frac12$. The stress tensor is 
\begin{equation}
T(z) \equiv \frac12 \omega_{ab} \no{\partial Z^a(z) Z^b(z)}
\end{equation}
It coincides with the Sugawara stress tensor for the $J^{ab}(z)$, simply because there is an unique $Sp(2n)$-invariant operator in dimension $2$. 

Basic correlation functions on a sphere can be computed from Wick contractions. The two-point function is 
\begin{equation}
\langle Z^a(z) Z^b(w) \rangle = \frac{\omega^{ab}}{z-w}
\end{equation}
etcetera. 

The theory of chiral symplectic bosons has subtle anomalies, which are ultimately motivated by the lack of a natural integration cycle for the 
path integral, which can at best be treated as a contour path integral. The choice of integration contour may depend on the geometry of 
space-time and on the choice of operator insertions, making correlation functions multi-valued: it is a relative CFT.\footnote{It is relative to a 3d TFT obtained by twisting a theory of free hypermultiplets \cite{Costello:2018fnz}.}

One may also consider a non-chiral version of the theory, combining the chiral symplectic bosons $Z^a$ 
with complex conjugate fields $\bar Z^{\bar a}$ with an action 
\begin{equation} \label{eq:ncbare}
\frac{1}{2 \pi} \int d^2 z \left[\omega_{ab} Z^a \bar\partial Z^b - \bar \omega_{\bar a \bar b} \bar Z^{\bar a} \partial \bar Z^{\bar b}\right]
\end{equation}
This combination has a natural integration contour where $\bar Z^{\bar a}$ is conjugate to $Z^a$ so that the action is pure imaginary. 
This is a true 2d CFT, with single-valued correlation functions. 

We will include a GSO projection in the definition of the non-chiral theory, restricting local operators to contain an even number of fields of half-integral spin 
and allowing vertex operators from both the Neveu-Schwarz and the Ramond sectors of the theory. This will be useful for the definition of minimal Hecke operators, which 
belong to the Ramond sector. The GSO-projected theory has $PSp(2n)$ global symmetry, as the action of the center of $Sp(2n)$ matches the spin parity of the fields.

\subsection{Vertex operators in the Neveu-Schwarz sector}
The Neveu-Schwarz sector consists of vertex operators around which the symplectic bosons are single-valued,
i.e. of vertex operators which can be placed at points around which the spin structure is bounding. Correspondingly, the Fourier modes 
\begin{equation}
Z^a_n \circ V(z) = \oint_z \frac{dw}{2 \pi i} w^{n-\frac12} Z(w) V(z)
\end{equation}
of the $Z^a$ fields have half-integral mode index $n$. Vertex operators which are polynomial in the $Z^a$ fields and their derivatives form the vacuum module of the chiral algebra. 

It is possible and often useful to define vertex operators which do not belong to the vacuum module. 
The calculation of correlation functions in the presence of such general vertex operators can be rather cumbersome.
A classical example of non-trivial vertex operators for a $\beta \gamma$ system occurs in the theory of superstrings: 
the $\delta(\beta)$ and $\delta(\gamma)$ vertex operators. By definition, these vertex operators impose a zero at their location for $\beta$ or $\gamma$ respectively, while allowing a pole for the other field.  In our setting, where $\beta$ and $\gamma$ appear on an even footing, we will need to consider generalizations such as such as $\delta(\beta + v \gamma)$. 

The standard bosonization technology employed in string theory textbooks is inadequate to deal with such a generalization. In the following we will employ an alternative perspective and computational strategy based on the idea that these vertex operators can be written as an integral transform of vertex operators which do belong to the vacuum module. 
A crucial aspect of this strategy is that the ``integral'' in the integral transform should be understood in a formal sense until a relatively late stage in the calculation: the specification of an integration contour is not part of the definition of the vertex operator, but rather part of the data which specifies a choice of conformal block. A useful perspective is that both conformal blocks in chiral algebras and integration contours in finite-dimensional integrals should be thought as a way to choose a solution of Ward identities for the respective systems. When the systems are combined, so are the notions of conformal block and integration contour.\footnote{A reader of a more algebraic bent may want to replace the formal integrals with constructions based on the theory of D-modules. The integrand will be expressed as a D-module which is also a chiral algebra module, and the integral is a push-forward operation.}

We use the integral transform
\begin{equation}
\delta(\beta + v \gamma) \simeq \oint d\lambda \, e^{\lambda (\beta + v \gamma)}
\end{equation} 
to map these vertex operators to exponential operators whose correlation function is readily computed. The correlation function may be multi-valued or 
have an intricate exponential behaviour at large $\lambda$. A choice of contour adapted to the situation will give access to a choice of conformal block for 
the correlation function with $\delta(\beta + v \gamma)$ insertion(s).

More generally, consider a collection of symplectic bosons $Z^a$ and vertex operators 
${\cal V}_v(z)$ which force some linear combination $v_a Z^a$ to vanish at $z$. In the chiral theory, we can write  
\begin{equation}
{\cal V}_v \equiv \; \oint d\lambda \, e^{\lambda v_a Z^a}\; \simeq  \;\delta\left(v_a Z^a\right)
\end{equation}

In the non-chiral theory, we will write
\begin{equation}
{\mathbb V}_v \equiv \int_{\mathbb{C}} d^{2} \lambda \, e^{\lambda v_a Z^a - \bar \lambda \bar v_{\bar a} \bar Z^{\bar a}}\; \simeq\;
 \delta^{(2)}\left(v_a Z^a, \bar v_{\bar a} \bar Z^{\bar a}\right)\end{equation}
Here we also do the integral after the correlation function has been computed, 
but we employ a canonical integration contour $\bar \lambda = \lambda^*$ over the complex $\lambda$ plane. 

We can now compute a sphere correlation function of such vertex operators. The correlation function of the exponential vertex operators is readily computed and we are left with a Gaussian integral
\begin{equation}
\left \langle \prod_{i=1}^k {\cal V}_{v^i}(z_i) \right\rangle = \int \prod_i d\lambda_i \, e^{\lambda_i \lambda_j M^{ij}[v;z]} = \frac{1}{\sqrt{\det M^{ij}[v;z]}}
\end{equation}
with an off-diagonal symmetric matrix 
\begin{equation}
M^{ij}[v;z] = \frac{\omega^{ab} v_a^i v_b^j}{z_i - z_j} \qquad \qquad i \neq j
\end{equation}

In the non-chiral theory, we have 
\begin{equation} \label{eq:corr}
\left \langle \prod_{i=1}^k {\mathbb V}_{v^i}(z_i, \bar z_i) \right\rangle = \int \prod_i d^2\lambda_i \, e^{\lambda_i \lambda_j M^{ij}[v;z]- \mathrm{c.c.}} = \frac{1}{|\det M^{ij}[v;z]|}
\end{equation}

Notice the divergence whenever $M^{ij}[v;z]$ has null eigenvectors. This is due to the existence of zeromodes for the $Z^a$ fields. 
Indeed, a classical vev for $Z^a$ would take the form 
\begin{equation}
Z^a_{cl}(z) = \sum_i \frac{\lambda_i \omega^{ab}v^i_b}{z-z_i}
\end{equation}
with the constraint that the contraction with $v^j_a$ of the finite part at $z_j$ 
\begin{equation}
\sum_{i\neq j} \frac{\lambda_i v^j_a \omega^{ab}v^i_b}{z_j-z_i}
\end{equation}
vanishes for all $j$. This is precisely the condition for the $\lambda_i$ to give a null eigenvector for $M^{ij}[v;z]$. 

The vertex operator ${\cal V}_{v}(z)$ has the following useful properties:
\begin{itemize}
\item It depends only on the direction of $v$, up to rescaling:  ${\cal V}_{v}= c {\cal V}_{c v}$. If we identify $v$ with the homogeneous coordinate 
of a point in a $\mathbb{CP}^{2n-1}$ parameter space, ${\cal V}_{v}$ behaves as a section of the ${\cal O}(-1)$ line bundle over the parameter space. 
\item It is a Virasoro primary of dimension $-\frac12$. 
\item It is {\it not} a primary for the Kac-Moody currents, as it is not annihilated by $J_1^{ab}$. We will momentarily discuss situations 
where ${\cal V}_{v}(z)$ is a primary for some Kac-Moody sub-algebra.
\end{itemize}

For completeness, we can describe explicitly the chiral algebra module generated by ${\cal V}_v$. The action of the $Z^a_m$ modes will 
pull down powers of $\lambda$ in the auxiliary integral, corresponding to vertex operators of the form $\delta'\left(v_a Z^a\right)$, etc. 
Denote the corresponding collection of states as $|v;m\rangle$, with $|v;0\rangle$ being the state corresponding to ${\cal V}_v$.
Then we have
\begin{align}
Z_n^a |v;m\rangle &= 0 \qquad \qquad \qquad \qquad m>\frac12 \cr
Z_{\frac12}^a |v;m\rangle &= \omega^{ab} v_b |v;m+1\rangle \cr
v_a Z_{-\frac12}^a |v;n\rangle &= - m |v;m-1\rangle
\end{align}
All other negative modes act freely. 

We also have a parallel transport along the $v^a$ parameter space:
\begin{equation}
\partial_{v_a} |v;m\rangle = Z_{-\frac12}^a |v;m+1\rangle
\end{equation}
This equation can be understood as holding when the corresponding vertex operators are inserted in correlation functions, or can be taken as the definition of a D-module structure on a bundle of chiral algebra modules over $\mathbb{CP}^{2n-1}$. 

\subsection{Sub-algebra primaries}
In applications, we will typically be interested in symplectic bosons which transform in a symplectic representation $R$ of some reductive group $G$. 
Correspondingly, we will only be interested in the $\mathfrak{g}$ Kac-Moody currents associated to the $G$ action. 
Concretely, we can pick generators $T^I_{ab}$ for the Lie algebra of $G$ and build the corresponding $\mathfrak{g}$ Kac-Moody currents 
\begin{equation}
J^{I}(z) \equiv \, \frac12 T^I_{ab} \no{Z^a(z) Z^b(z)}
\end{equation}
The level of the currents is proportional to the second Casimir of $R$. 

In this situation, we can reassess the properties of the vertex operator ${\cal V}_{v}(z)$. The action of the $J_1^I$ currents on ${\cal V}_{v}(z)$ now involves
the moment map $v_a \to T^{I;ab} v_a v_b$. If the representation admits ``pure'' vectors $v_a$ such that $T^{I;ab} v_a v_b$ vanishes for all $I$, 
we can use them to define highest weight vertex operators for the $\mathfrak{g}$ Kac-Moody algebra.

Assume now that $v_a$ is a pure vector. We can readily compute the action on ${\cal V}_{v}(z)$ of the $J_{0}^I$ zeromodes. We have 
\begin{equation}
J^I_0 |v;0\rangle = T^I_{ab} Z_{-\frac12}^a Z^b_{\frac12} |v;0\rangle = T^I_{ab} \omega^{bc} v_c Z_{-\frac12}^a |v;1\rangle 
\end{equation}
i.e.
\begin{equation}
J^I_0 |v;0\rangle = T^I_{ab} \omega^{bc} v_c \partial_{v_a} |v;0\rangle 
\end{equation}
In other words, the zeromodes act as differential operators on $v$, implementing the action of the Lie algebra of $G$
on the space of pure vectors.  

As the vertex operators are co-variant under re-scaling of $v$, the action corresponds to the action of the Lie algebra on sections of a certain line bundle on the projective space of pure vectors. 

The insertion of $\mathfrak{g}$ currents in a sphere correlation function can be expressed as a differential operator. For example, 
\begin{equation}
\left \langle J^I(z) \prod_{i=1}^k {\mathbb V}_{v^i}(z_i, \bar z_i) \right\rangle = \left[\sum_i \frac{1}{z-z_i} T^I_{ab} \omega^{bc} v^i_c \partial_{v^i_a} \right] \frac{1}{|\det M^{ij}[v;z]|}
\end{equation}

There are two particularly interesting situations, the second of which is the one relevant to this paper:
\begin{itemize}
\item If the level of the $\mathfrak{g}$ currents is not critical, we may have a {\it conformal embedding}, i.e. the stress tensor of the theory may coincide with the Sugawara stress tensor for the $\mathfrak{g}$ currents. Then the correlation function (\ref{eq:corr}) will be a solution of Knizhnik–Zamolodchikov differential equations, equating the $z_i$ derivative with a second order differential operator in the $v_j$'s. 
\item If the level of the $\mathfrak{g}$ currents is critical, we obtain a map from the center $Z[\widehat{\mathfrak{g}}_{-h^\vee}]$ of the universal enveloping algebra $U[\widehat{\mathfrak{g}}_{-h^\vee}]$ of the critical $\mathfrak{g}$ Kac-Moody algebra 
to the chiral algebra of the symplectic bosons. This map will often not be injective, leading to linear relations between the images of operators in $Z[\widehat{\mathfrak{g}}_{-h^\vee}]$. The insertion of operators in $Z[\widehat{\mathfrak{g}}_{-h^\vee}]$ in a correlation function can be expressed as the action of certain differential operators 
in the $v_i$'s. Any linear relation between the images will become a non-trivial differential equation satisfied by the correlation function (\ref{eq:corr}).
\end{itemize}

\subsection{Weighed powers}
The construction of the $\delta$-function vertex operators is just one example of a vast zoo of possibilities. A simple but perhaps surprising modification is the following: 
\begin{equation}
{\cal V}_{v;\alpha} \; := \oint d\lambda \, \lambda^{-\alpha-1} {e^{\lambda v_a Z^a}}
\end{equation}
which morally speaking represents a fractional power ${(v_a Z^a)^\alpha}$ of an elementary field. The module associated to such a vertex operator includes vertex operators with all values of $\alpha$ which differ by integer values. 

The fractional power of $\lambda$ in the integrand leads to a greater variety of possible integration contours. For example, a sphere correlation function becomes 
 \begin{equation}
\left \langle \prod_{i=1}^k {\cal V}_{v^i,\alpha_i}(z_i) \right\rangle =\oint \left[\prod_i d\lambda_i \,\lambda_i^{-\alpha_i-1} \right]\, e^{\lambda_i \lambda_j M^{ij}[v;z]}
\end{equation}

In the non-chiral theory, we write an integral of the form 
\begin{equation}
{\mathbb V}_{v;\alpha} \; := \int_{\mathbb{C}} d^{2} \lambda \, \lambda^{-\alpha-1} \bar \lambda^{-\tilde \alpha-1}\no{e^{\lambda v_a Z^a - \bar \lambda \bar v_{\bar a} \bar Z^{\bar a}}}
\end{equation}
with constraints $\alpha - \tilde \alpha \in \mathbb{Z}$ for single-valuedness and $\mathrm{Re} [\alpha + \tilde \alpha]=0$ for (conditional) convergence. 

The non-chiral correlation function becomes
 \begin{equation}
\left \langle \prod_{i=1}^k {\mathbb V}_{v^i,\alpha_i}(z_i) \right\rangle =\int \left[\prod_i d^2 \lambda_i\, \lambda_i^{-\alpha_i-1} \bar \lambda_i^{-\tilde \alpha_i-1} \right]\, e^{\lambda_i \lambda_j M^{ij}[v;z]- \bar \lambda_i \bar \lambda_j M^{ij}[\bar v;\bar z]}
\end{equation}

These modified vertex operators for a pure $v$ will still be Kac-Moody primaries, with the current zeromodes acting as differential operators on $v$, implementing the action of the Lie algebra of $G$
on the space of pure vectors. The main effect of the weight $\alpha$ is to change the behaviour of the vertex operator under scaling of $v$, so that the action of the Lie algebra zero-modes on the projective space of pure vectors is twisted by the $\alpha$-th power of the tautological line bundle.

\subsection{Ramond vertex operators}
Vertex operators in the Ramond sector introduce a $Z^a \to - Z^a$ cut starting from the vertex operator location. 
These vertex operators should thus always appear in pairs. The $Z^a_n$ Fourier modes acting on a Ramond vertex operator have integral mode index $n$. 
We will focus on highest weight Ramond vertex operators, which are annihilated by all the modes with positive $n$. 

Highest weight Ramond vertex operators form modules for the Weyl algebra of zeromodes
\begin{equation}
[Z^a_0, Z^b_0] = \omega^{ab}
\end{equation}
Conversely, any such Weyl module can be promoted to a collection of highest weight Ramond vertex operators related by the symplectic boson Ward identities. 

In the chiral theory there is no canonical choice of Weyl module module, though vertex operators belonging to different modules can often be related to each other by 
formal integral transformations. 

The non-chiral theory has holomorphic and anti-holomorphic copies of the zeromode algebra
\begin{equation}
[Z^a_0, Z^b_0] = \omega^{ab} \qquad \qquad [\bar Z^{\bar a}_0, \bar Z^{\bar b}_0] = -\bar \omega^{\bar a \bar b}
\end{equation}
and we can look for representations where $Z^a_0$ are hermitean conjugate to $\bar Z^a_0$ and both are normal. 
The natural choice is to employ the standard representation on the 
Hilbert space $L^2(\mathbb{C}^{n})$. Concretely, we can pick a Lagrangian splitting of $\mathbb{C}^{2n}$ and have half of the chiral modes act as holomorphic multiplication operators and the other half as holomorphic derivatives. The anti-chiral modes can act by anti-holomorphic differential operators. 

Different choices of polarization are related as usual by generalized Fourier transformations, so this prescription for the Ramond sector of the non-chiral theory is 
canonical. Equivalently, the space of states in the Ramond sector of the non-chiral theory is built from the space of states of the zero-mode quantum mechanics with $\mathbb{C}^{2n}$ phase space by acting freely with the negative modes of the algebra. Working with distributional states, we can realize a variety of modules for the chiral algebra. 

In the chiral theory we can still define Ramond modules by a Lagrangian splitting, starting from a highest weight vertex operator annihilated by half of the zeromodes. 
Vertex operators for different Lagrangian splittings can be formally related by a contour version of the generalized Fourier transformations. The choice of contours makes correlation functions multi-valued.

As an example, consider a single copy of the $\beta\gamma$ system. We can define an highest weight Ramond vertex operator ${\cal R}_0$ annihilated by $\gamma_0$. Algebraically, the rest of the Weyl module is generated by monomials $\beta_0^k \circ {\cal R}_0$. A formal linear combination 
$e^{\alpha \beta_0} \circ {\cal R}_0$ is annihilated by $\gamma_0 + \alpha$. The formal integral 
\begin{equation}
\int d \alpha \left[e^{\alpha \beta_0} \circ {\cal R}\right] 
\end{equation}
gives a vertex operator annihilated by $\beta_0$. 

In the non-chiral theory we consider Ramond vertex operators $R[f]$ labelled by elements $f(\beta_0, \bar \beta_0)$ in $L^2(\mathbb{C})$. 
The analogue of ${\cal R}_0$ is a (``non-normalizable'') vertex operator $R_0\equiv R[1]$ labelled by the distribution ``$1$''. A vertex operator 
$R[\delta^{(2)}(\beta_0, \bar \beta_0)]$ would be annihilated by $\beta_0$ and $\bar \beta_0$, etc. 

Highest weight Ramond modules are also highest weight modules for the Kac-Moody currents. The current zeromodes act on the highest weight vectors by the obvious bilinears $Z^{(a}_0 Z^{b)}_0$, giving the metaplectic action of the group. 

\subsection{An example of Ramond correlation functions}
We can elaborate on the example of a single pair of symplectic bosons. We will denote as ${\cal R}_v$ highest weight Ramond vertex operators annihilated by $\gamma_0 + v \beta_0$. Here $v$ should be thought of as an inhomogeneous coordinate on a $\mathbb{CP}^1$ family of vertex operators. 

We can adjust the normalization of these operators in such a way that $SL(2)$ rotations act on them as (twisted) differential operators acting on $v$. The metaplectic action on the vertex operators has an anomaly, which results in the vertex operators transforming as sections of the twisted line bundle ${\cal O}(-\frac12)$ over 
$\mathbb{CP}^1$. Concretely, under 
\begin{equation}
\beta \to a \beta + c \gamma \qquad \qquad  \gamma \to b \beta + d  \gamma
\end{equation}
we have 
\begin{equation}
\gamma_0 + v \beta_0 \to  (c v + d) \gamma_0 + (a v + b) \beta_0
\end{equation}
but 
\begin{equation}
{\cal R}_{v} = (c v + d)^{-\frac12} {\cal R}_{\frac{a v+b}{c v + d}}
\end{equation}
We can verify this by writing formally 
\begin{equation}
{\cal R}_{v} = e^{\frac{v}{2} \beta_0^2} \circ {\cal R}_{0}
\end{equation}
and looking at the effect of an inversion $v \to -v^{-1}$, which is implemented by a formal Fourier transform of the Gaussian. 

We can get some immediate insight on a sphere correlation function 
\begin{equation}
\left\langle \prod_{i=1}^{2k} {\cal R}_{v_i}(z_i) \right\rangle
\end{equation}
by looking for values of the $v_i$ which admit zeromodes for $\beta$ and $\gamma$. A zeromode appears if we have a classical solution with appropriate boundary conditions. 
Concretely, we can write classical vevs
\begin{equation}
\beta(z) = \frac{B_{k-1}(z)}{\sqrt{\prod_i (z-z_i)}} \qquad \qquad \gamma(z)= \frac{C_{k-1}(z)}{\sqrt{\prod_i (z-z_i)}}
\end{equation}
where $B_{k-1}$ and $C_{k-1}$ are degree $k-1$ polynomials. The degree is determined by regularity at infinity, where a spin $\frac12$ field must 
go to $0$ as $z^{-1}$.

The zeromode constraint at the Ramond punctures requires 
\begin{equation}
C_{k-1}(z_i)+ v_i B_{k-1}(z_i)=0
\end{equation}
for all $i$. These are $2k$ linear equations in $2k$ linear variables, the coefficients of the two polynomials. 

We can solve the system of linear equations iff $\det P(v_*;z_*) =0$, where $P(v_*;z_*)$ is the $2k \times 2k$ matrix with entries $z_i^{j-1}$ for $j\leq k$ and $v_i z_i^{j-1-k}$ for $j>k$. The appearance of a zeromode causes an inverse square root divergence in the correlation function. Inspired by that, we 
propose the tentative ansatz
\begin{equation} \label{eq:example}
\left\langle \prod_{i=1}^{2k} {\cal R}_{v_i}(z_i) \right\rangle = \frac{ \prod_{i<j} (z_i - z_j)^{\frac14} }{\sqrt{\det P}}
\end{equation}
This has the correct divergence when zeromodes appear and has the correct behaviour under fractional linear transformations of the $v_i$. The $z_i$-dependent numerator also insures the correct behaviour under global conformal transformations, i.e. fractional linear transformations of the $z_i$, 
compatibly with the scaling dimension $-\frac18$ of the ${\cal R}_{v}$'s. 

We will now verify the correctness of this ansatz with the help of spectral flow Ward identities. 

\subsection{Spectral flow}
The term spectral flow denotes a family of chiral algebra automorphisms which generalize of the inner automorphisms induced by the modes of a Kac-Moody sub-algebra. 

The spectral flow automorphisms of the symplectic boson VOA are particularly simple to describe. Consider a locally holomorphic, position-dependent $Sp(2n)$ group element $g(z)$. The singular part of the OPE of symplectic bosons is invariant under position-dependent holomorphic symplectic rotations
\begin{equation}
Z^a(z) \to g^a_b(z) Z^b(z)
\end{equation}
as we can Taylor expand
\begin{equation}
\frac{g^a_b(z) g^{a'}_{b'}(w)\omega^{b b'}}{z-w} \sim \frac{\omega^{a a'}}{z-w} + \cdots
\end{equation}
This symmetry has both local and global consequences. 

Locally, we can take $g(z)$ to be defined on a small annular region around some point $z=0$. Then 
the spectral flow automorphism maps 
\begin{equation}
Z^a_n \to \oint_z \frac{dw}{2 \pi i} w^{n-\frac12} g^a_b(w) Z^b(w)
\end{equation}
and preserves the mode algebra. Given any module for the mode algebra, we can map it to a new spectral-flowed module depending on the choice of $g(z)$. 

The spectral flow automorphism depends on the Laurent expansion of $g(z)$ at $w$. The space of such Laurent expansions is 
usually denoted as $Sp(2n)[{\cal K}]$. The space of regular Taylor expansions is usually denoted as $Sp(2n)[{\cal O}]$. As the vacuum vector itself is invariant under $Sp(2n)[{\cal O}]$ transformations, the image of the vacuum under the spectral flow automorphism only depend on the class $[g]$ of $g$ in 
$Sp(2n)[{\cal K}]/Sp(2n)[{\cal O}] \equiv \mathrm{Gr}_{Sp(2n)}$, i.e. the {\it affine Grassmanian} of the symplectic group. 

If we take a multivalued group transformation which satisfies $g(e^{2 \pi i} z) = - g(z)$, the spectral flow automorphism will relate Neveu-Schwarz and Ramond modules. 
If we pick a Lagrangian splitting of the $Z^a$ and take $g$ to act as $z^{\pm \frac12}$ on the two subspaces, the image of the vacuum under $g$ will be precisely 
the highest weight Ramond vertex operators annihilated by the corresponding half of the $Z^a_0$ zeromodes, and viceversa. 

If we are given a globally defined meromorphic $g(z)$ on a Riemann surface, we can employ it to relate different correlation functions. 
Indeed, a correlation function of transformed fields $g^a_b(z) Z^b(z)$ will satisfy the same Ward identities 
as a correlation function of $Z^a(z)$, except at points where the symplectic rotation $g(z)$ becomes singular. 
At those points we need to apply the spectral flow automorphism defined locally by $g(z)$. We will call such a relation 
a ``spectral flow Ward identity''. 

An important caveat for this statement is that it is most useful if we know how {\it all} vertex operators in the correlation function 
transform as $Z^a(z) \to g^a_b(z) Z^b(z)$. At regular points $z_i$ of $g(z)$, the transformation can be computed as the exponential of 
the action of the non-negative Kac-Moody algebra modes. If we only have highest 
weight vertex operators for the Kac-Moody algebra, only the Kac-Moody zeromodes act non-trivially and 
we just apply the constant rotation $g(z_i)$ to the vertex operator.  

A simplifying feature is that spectral flow operations compose well, so that if we transform by $g(z)$ the image of a vertex operator under some other $g'(z)$, 
we get the image under $g(z) g'(z)$, up to a computable phase anomaly proportional to the level of the Kac-Moody currents. 

As a simple example, consider again the case of a single $\beta \gamma$ system. We can employ the global spectral flow operation on the sphere
\begin{equation}
\beta(z) \to z^{\frac12} \beta(z) \qquad \qquad \gamma(z) \to z^{-\frac12} \gamma(z)
\end{equation}
to add highest weight Ramond vertex operators at $0$ and $\infty$: 
\begin{equation}
\left \langle \prod_{i=1}^{2k} {\cal R}_{v_i}(z_i)\right \rangle = \left \langle {\cal R}_{\infty}(\infty) \left[\prod_{i=1}^{2k} z_i^{\frac14} {\cal R}_{z_i v_i}(z_i) \right]{\cal R}_{0}(0)\right \rangle
\end{equation}
This spectral flow Ward identity, combined with the $SL(2)$ global symmetries acting on the $v_i$ and on the $z_i$, determines the correlation functions recursively. It is satisfied by our ansatz (\ref{eq:example}).

\subsection{Boundary conditions}
We have given tools to study the non-chiral version of the symplectic boson theory on an orientable Riemann surface without boundaries. With some extra structure, the non-chiral theory can also be placed on Riemann surfaces which have a boundary and/or are unorientable.  

In this paper we will only consider correlation functions on a disk or cross-cap. We will also focus on boundary conditions or cross-cap conditions which allow a doubling trick, 
so that the correlation function of the non-chiral theory becomes a correlation function for the chiral theory on a sphere, with operator insertions at $z$ and $\bar z$ or at $z$ and $-\bar z^{-1}$ which represent the chiral and anti-chiral halves of the non-chiral vertex operators. 

Because of the quadratic nature of the action and spinorial nature of the $Z^a$'s, boundary conditions and cross-caps conditions will typically occur in pairs, 
related by $Z^a \to Z^a$ together with $\bar Z^a \to - \bar Z^a$. Our final prescription will be the sum of correlation functions for the two choices,
which has better reality properties. 

For example, the simplest possible boundary conditions  for symplectic bosons impose
\begin{equation}
\bar Z^a(z) = \pm i Z^a(z)
\end{equation}
along the real $z$ axis. The imaginary prefactor follows from our conventions where the chiral and anti-chiral actions and OPE have opposite signs. 

These boundary conditions support boundary vertex operators with properties analogous to these of NS sector chiral vertex operators. Boundary-changing vertex operators
which interpolate between these boundary conditions have properties analogous to these of R sector chiral vertex operators. Both types of vertex operators will be supported on the 
direct sum of the two boundary conditions. 

As an example of a disk correlation functions with bulk insertions only, consider 
\begin{equation} \label{eq:corr}
\left \langle \prod_{i=1}^k {\mathbb V}_{v^i}(z_i, \bar z_i) \right\rangle_{\mathbb{D},\pm} = \int \prod_{i=1}^{2k} d\tilde \lambda_i \, e^{\tilde \lambda_i \tilde \lambda_j M^{ij}[\tilde v;\tilde z]} 
\end{equation}
where we collected into $2n$-component vectors the combined data of the chiral and anti-chiral halves: 
\begin{align}
\tilde z_{2j-1} &= z_j \qquad \qquad 
\tilde z_{2 j} = \bar z_j \cr 
\tilde v_{2j-1} &= v_j \qquad \qquad 
\tilde v_{2 j} = \bar v_j \cr 
\tilde \lambda_{2j-1} &= \lambda_j  \qquad \qquad 
\tilde \lambda_{2 j} = \pm i \bar \lambda_j 
\end{align}
The sign is determined by the choice of boundary condition. 

Define $\sigma$ as the matrix which permutes the $2i$ and $2i-1$ entries of a $2k$-dimensional vector for all $i$. Then 
the matrix $M$ satisfies $M^* = \sigma M \sigma$ while the vector $\tilde \lambda$ satisfies $\sigma (\tilde \lambda)^* = \mp i \tilde \lambda$. 
For both signs, the reality condition on $\tilde \lambda$ guarantees that the exponent in the Gaussian integral is pure imaginary. 
The integral should give a branch of the function $\frac{1}{\sqrt{\det M}}$. 

The two answers will thus either coincide or have opposite values.
It is not hard to check that the two answers must be conjugate to each other, say by diagonalizing $\sigma$ to get a multi-dimensional version of our Gaussian example. 
Their sum then gives the real, possibly vanishing, answer $\frac{\theta(\det M)}{\sqrt{\det M}}$. 

The disk correlation function could be enriched by boundary vertex operators ${\mathbb V}^\partial_{v}(z)$, only defined for real $z$. We can preserve the nice reality properties
of the answer if we also restrict $v$ to be real and integrate the corresponding $\lambda$ along $\lambda^* = \mp i \lambda$. 

The basic boundary conditions for the disk can be modified to 
\begin{equation}
\bar Z^a(z) = \pm i g^a_b Z^b(z)
\end{equation}
for some element $g$ in $Sp(2n,\mathbb{C})$. In order to have correlation functions with good reality properties, we can restrict to $g$ which satisfy $g g^* =1$. 
Then, for example, $\tilde v_{2 j} = \bar v_j g$ is still compatible with $\sigma$ being an involution. Boundary vertex operators with good reality conditions 
will satisfy $v = \bar v g$. Depending on the choice of $g$, they may not exist. 

For cross-cap correlation functions, we would have $\tilde z_{2 j} = - \bar z_j^{-1}$. The analysis can proceed roughly as before, with $\sigma$ involving multiplication by $z_j^{\pm 1}$. We leave the details to an enthusiastic reader. 

\section{Tri-fundamental symplectic bosons} \label{sec:trifu}
In this section we specialize to the case of symplectic bosons valued in $\mathbb{C}^8 = \mathbb{C}^2 \otimes\mathbb{C}^2 \otimes\mathbb{C}^2$
with an $SL(2) \times SL(2)\times SL(2)$-invariant symplectic form. Accordingly, we denote the symplectic boson fields as $Z^{\alpha \beta \gamma}(z)$ and the symplectic form as
$\epsilon_{\alpha \alpha'}\epsilon_{\beta \beta'}\epsilon_{\gamma \gamma'}$. After GSO projection, the non-chiral theory will have $\mathrm{PSL}(2) \times \mathrm{PSL}(2)\times \mathrm{PSL}(2)$
global symmetry.

The corresponding chiral algebra $V_{\mathfrak{sl}_2,3}$ includes three sets of $\widehat{\mathfrak{sl}}(2)_{-2}$ Kac-Moody currents such as 
\begin{equation}
J^{\alpha \alpha'} =\frac12 \epsilon_{\beta \beta'}\epsilon_{\gamma \gamma'} :Z^{\alpha \beta \gamma}Z^{\alpha' \beta' \gamma'}:
\end{equation}
These currents have critical level. Accordingly, the Sugawara vectors are central. Remarkably, they coincide:
\begin{equation}
:\det J: = :\det J': = :\det J'':
\end{equation}
This can be demonstrated by a simple counting argument: there are only two $SL(2)^3$-invariant descendants in the vacuum module at level 2, and 
one of them is the actual stress tensor of $V_{\mathfrak{sl}_2,3}$, which is not central.  

We will now define several special vertex operators:
\begin{itemize}
\item Neveu-Schwarz vertex operators ${\cal V}(x,x',x'';z)$ which are highest weight for the three $\widehat{\mathfrak{sl}}(2)_{-2}$'s Kac-Moody algebras,
with zeromodes acting as in (\ref{eq:basic}). Sphere correlation functions of ${\cal V}(x,x',x'';z)$ give the chiral kernels $K_3$. 
\item Deformed Neveu-Schwarz vertex operators ${\cal V}_\alpha(x,x',x'';z)$ which are highest weight for the three $\widehat{\mathfrak{sl}}(2)_{-2}$'s Kac-Moody algebras with a more general weight. Sphere correlation functions of ${\cal V}_\alpha(x,x',x'';z)$ give the general weight variants of $K_3$. We will not employ them in concrete calculations below, but it is not difficult to do so. 
\item Ramond vertex operators ${\cal W}(x,x',x'';z)$ which are also highest weight for the three $\widehat{\mathfrak{sl}}(2)_{-2}$'s with the zeromodes acting as in (\ref{eq:basic}). These are related to ${\cal V}(x,x',x'';z)$ by a spectral flow automorphism of either of the three $\mathrm{PSL}(2)$'s. Replacing 
${\cal V}(x,x',x'';z)$ with ${\cal W}(x,x',x'';z)$ (or viceversa) in a correlation function is the same as acting with a special Hecke operator at $z$ for any of the three $\mathrm{PSL}(2)$'s. 
\item A Ramond vertex operator ${\cal M}(z)$ which has trivial OPE with the $\widehat{\mathfrak{sl}}(2)_{-2}$'s Kac-Moody currents. 
This is related to the vacuum vector by an {\it averaged} spectral flow automorphism for either of the three $\mathrm{PSL}(2)$'s. Inserting ${\cal M}(z)$ in a correlation function 
is the same as acting with an integral Hecke operator at $z$ for any of the three $\mathrm{PSL}(2)$'s.
\end{itemize} 

The very existence of these vertex operators guarantees that the non-chiral kernels ${\mathbb K}_3$ and their weighed variants will intertwine 
the corresponding Hecke operators as well as the Gaudin Hamiltonians. 

\subsection{Genus zero correlation functions and parabolic bundles}
Our first example of vertex operator is the Neveu-Schwarz delta-function vertex operator  ${\cal V}_{v}(z)$ with 
a pure vector 
\begin{equation}
v_{\alpha \beta \gamma} = x_\alpha x'_\beta x''_\gamma
\end{equation}
in $\mathbb{C}^8$. We denote this specialization as ${\cal V}(x,x',x'';z)$. 

The vertex operator is covariant under a rescaling of $x$, $x'$ or $x''$, 
so this is a $\mathbb{CP}^{1} \times\mathbb{CP}^{1} \times\mathbb{CP}^{1}$ family of vertex operators, transforming as a section of 
${\mathcal O}(-1,-1,-1)$. It is often useful to pick inhomogeneous coordinates $x_{\alpha} = (1,x)$, etc. 

As discussed in the previous section, these are highest weight vectors for 
the three sets of $\widehat{\mathfrak{sl}}(2)_{-2}$ Kac-Moody currents. 
We have an OPE
\begin{equation}
J(z) {\cal V}(x,x',x'';0) \sim \frac{{\cal L}\circ {\cal V}(x,x',x'';0)}{z} + \cdots
\end{equation}
where 
\begin{equation}
{\cal L}^{\alpha \beta} = \frac12 \epsilon^{\alpha \gamma} x_\gamma \partial_{x_\beta} +  \frac12 \epsilon^{\beta \gamma} x_\gamma \partial_{x_\gamma}
\end{equation}
are $\mathfrak{sl}(2)$ generators acting as vector fields on sections of $O(1)$ over $\mathbb{C}P^1$. That is the same as (\ref{eq:basic}) when written in
inhomogeneous coordinates.

The genus $0$ correlation function defines the chiral kernel 
\begin{equation}
K_3(x_*,x_*',x_*'';z_*) \equiv \left\langle \prod_{i=1}^n {\cal V}(x_i,x_i',x_i'';z_i) \right\rangle_{\mathbb{CP}^1} = \frac{1}{\sqrt{\det A}}
\end{equation}
computed from the matrix 
\begin{equation}
A^{ij} = \frac{(x_i,x_j)(x'_i,x'_j)(x''_i,x''_j) }{z_i-z_j} \qquad \qquad i \neq j
\end{equation}
and $A^{ii}=0$. Here we defined  
\begin{equation}
(x_i,x_j) \equiv \epsilon^{\alpha \beta} x_{i,\alpha}x_{j,\beta} = x_i - x_j
\end{equation}
etcetera. Notice that a rescaling of any homogeneous variable, say $x_{1,\alpha} \to \eta x_{1,\alpha}$, results in a rescaling 
of $\det A \to \eta^2 \det A$ and thus $K_3\to \eta^{-1} K_3$. Hence $K_3$ transforms as a section of ${\mathcal O}(-1)$, as it should.

The insertion of the Sugawara vector $:\det J:(z)$ into the correlation function results in the action of the differential operator 
\begin{equation}
H(z) = \sum_i \frac{H_i}{z-z_i}
\end{equation}
where $H_i$ are the Gaudin Hamiltonians 
\begin{equation}
H_i =  \sum_{j \neq i} \frac{{\cal L}_i \cdot {\cal L}_j}{z_i-z_j}
\end{equation}
built from the $\mathfrak{sl}(2)$ action on the $x$'s. We have checked explicitly for small $n$ that $K_3$ indeed intertwines the Gaudin Hamiltonians
acting on the three sets of variables.  

We can replace any ${\cal V}(x,x',x'';z)$ vertex operator with a weighed variant ${\cal V}_\alpha(x,x',x'';z) \equiv {\cal V}_{v;\alpha}(z)$ for 
$v_{\alpha \beta \gamma} = x_\alpha x'_\beta x''_\gamma$. The vertex operators now transform as a section of $O(\alpha, \alpha, \alpha)$ on 
the $\mathbb{CP}^{1} \times\mathbb{CP}^{1} \times\mathbb{CP}^{1}$ parameterized by $x$, $x'$, $x''$. 
This changes accordingly the highest weight of the puncture in the Gaudin Hamiltonians. 

\subsection{Translation to Hitchin Hamiltonians}
In order to make contact with the moduli space ${\cal M}_0(\mathrm{PSL}(2),\mathbb{CP}^1_n)$ of $\mathrm{PSL}(2)$ bundles of degree $0$, genus zero and $n$ parabolic points, observe that $K_3$ is covariant under simultaneous $SL(2)$ rotations of the $x_i$ variables. It can be identified with a (multivalued)
section of the square root of the canonical bundle over the space of $x_i$ modulo $SL(2)$ rotations, which is essentially ${\cal M}_0(\mathrm{PSL}(2),\mathbb{CP}^1_n)$. 

To be concrete, we can gauge-fix $SL(2)$ rotations by setting
$x_{n-2}$, $x_{n-1}$, $x_{n}$ to fixed values. The correlation function gives a wavefunction
\begin{equation}
K_3(x_*,x'_*,x''_*) \sqrt{d\mu(x_*) d\mu(x'_*) d\mu(x''_*)}
\end{equation}
where the gauge-fixed holomorphic top form $d\mu(x)$ on the space of parabolic bundles is
\begin{equation}
d\mu(x_*) = (x_{n-2},x_{n-1})(x_{n-1},x_{n})(x_{n},x_{n-2}) \prod_{i=1}^n dx_i
\end{equation}
and $dx$ is the holomorphic top form on $\mathbb{C}P^{1}$. The Gaudin Hamiltonians acting on such $SL(2)$-invariant wavefunctions become the quantum Hitchin Hamiltonians for ${\cal M}_0(SL(2),\mathbb{CP}^1_n)$. 

The non-chiral kernels $\mathbb{K}^{\C}_3$ defined by non-chiral sphere correlation functions give single-valued half-densities on ${\cal M}_0(\mathrm{PSL}(2),\mathbb{CP}^1_n)$ which intertwine both the holomorphic and anti-holomorphic Hitchin Hamiltonians and can be employed in the analytic Langlands program over the complex numbers. 
Disk or cross-cap correlation functions, with bulk and/or boundary vertex operator insertions, give half-densities $\mathbb{K}^{\mathbb{R}}_3$ on a variety of real versions ${\cal M}^{\mathbb{R}}_0(\mathrm{PSL}(2),\mathbb{CP}^1_n)$ of ${\cal M}_0(\mathrm{PSL}(2),\mathbb{CP}^1_n)$ which can be employed in the analytic Langlands program over the real numbers. 

\subsection{Ramond modules}
The tri-fundamental symplectic bosons admit some highest weight Ramond modules with very special properties. 
These descend from special modules for the Weyl algebra of $\mathbb{C}^8$, generated by the zeromodes $Z_0^{\alpha \beta \gamma}$.

The first module has the special property that the action of $SL(2) \times SL(2) \times SL(2)$ is integrable. It consists of an infinite direct sum 
\begin{equation}
\bigoplus_{n=1}^\infty \mathbb{C}^{n} \otimes \mathbb{C}^{n} \otimes \mathbb{C}^{n}
\end{equation}
where each factor in the product is the $n$-dimensional irreducible representation of the corresponding $SL(2)$. Denote a basis in each 
summand as $|M;\alpha_1 \cdots \alpha_n;\beta_1 \cdots \beta_n;\gamma_1 \cdots \gamma_n\rangle$, totally symmetric within each group of indices. 
Then the module structure is fixed by symmetry
\begin{align}
Z_0^{\alpha \beta \gamma} &|M;\alpha_1 \cdots \alpha_n;\beta_1 \cdots \beta_n;\gamma_1 \cdots \gamma_n\rangle = |M;\alpha\alpha_1 \cdots \alpha_n;\beta \beta_1 \cdots \beta_n;\gamma \gamma_1 \cdots \gamma_n\rangle + \cr &+ c_n\epsilon^{\alpha (\alpha_n}\epsilon^{\beta (\beta_n}\epsilon^{\gamma (\gamma_n} |M;\alpha_2 \cdots \alpha_n);\beta_2 \cdots \beta_n);\gamma_2 \cdots \gamma_n)\rangle
\end{align}
where the round parentheses denote symmetrization of the $\alpha_*$'s, $\beta_*$'s and $\gamma_*$'s and the numerical coefficients $c_n$ can be determined systematically by imposing the commutation relations. 

The cyclic vector $|M\rangle$ satisfies relations
\begin{equation}
	\epsilon_{\beta \beta'}\epsilon_{\gamma \gamma'}\left(Z^{\alpha \beta \gamma}_0Z^{ \alpha' \beta' \gamma'}_0+ Z^{\alpha' \beta \gamma}_0Z^{ \alpha \beta' \gamma'}_0\right) |M\rangle  = 0
\end{equation}
which simply encode $SL(2) \times SL(2) \times SL(2)$-invariance but are sufficient to determine recursively the full structure of the module. The existence of such a module is somewhat miraculous.  We can de-mistify by breaking one of the three $SL(2)$ symmetries and realizing the Weyl algebra as the algebra of polynomial differential operators in the $Z^{+\beta \gamma}_0$ variables. Starting from 
\begin{equation}
	\det_{\beta \gamma} Z^{+\beta \gamma}_0 |M\rangle =\epsilon_{\beta \beta'}\epsilon_{\gamma \gamma'}Z^{+\beta \gamma}_0Z^{+\beta' \gamma'}_0 |M\rangle  = 0
\end{equation}
we can attempt the identification 
\begin{equation}
	|M\rangle \simeq \delta\left(\det_{\beta \gamma} Z^{+\beta \gamma}_0\right)
\end{equation}
with a formal distribution. This is distribution is clearly invariant under the unbroken $SL(2)$'s. The condition 
\begin{equation}
	\epsilon_{\beta \beta'}\epsilon_{\gamma \gamma'}\left(Z^{+ \beta \gamma}_0Z^{- \beta' \gamma'}_0+ Z^{- \beta \gamma}_0Z^{+ \beta' \gamma'}_0\right) \delta\left(\det_{\beta'' \gamma''} Z^{+\beta'' \gamma''}_0\right)=0
\end{equation}
is less obvious but easy to check. The final condition 
\begin{equation}
	\left(\det_{\beta \gamma} Z^{-\beta \gamma}_0\right)  \delta\left(\det_{\beta' \gamma'} Z^{+\beta' \gamma'}_0\right)= 0
\end{equation}
is truly remarkable. 

We will denote the Ramond vertex operator corresponding to the 
cyclic vector $|M\rangle$ as ${\cal M}(z)$. By construction, it has non-singular OPE with the three sets of $\mathfrak{sl}(2)_{-2}$ Kac-Moody currents, i.e. it is central.

Starting from ${\cal M}(z)$ we can build a whole tower of central Ramond vertex operators: we can take $z$ derivatives and we can act with modes of the Sugawara vectors. 
A simple counting argument, though, shows that the module only includes two $SL(2)^3$-invariant vectors at level $2$. One of these is the Virasoro descendant $L_{-2} {\cal M}(z)$ 
which is not central. We thus must have a linear relation between two candidate central elements at level $2$: $\partial^2 {\cal M}(z)$ and the Sugawara descendant $S_{-2}{\cal M}(z)$.

The only possibility is that ${\cal M}(z)$ satisfies the classical limit of the BPZ differential equations, i.e. the oper equation:
\begin{equation}
\partial_z^2 {\cal M}(z) + S(z) {\cal M}(z) 
\end{equation}
with the Sugawara vector playing the role of the classical stress tensor. 

This property agrees with an important observation from \cite{Gaiotto:2021tsq}: the insertion of (the non-chiral version of) ${\cal M}(z)$ in a correlation function represents 
the action of any of the three minimal Hecke operators on the correlation function, seen as a half-density on the appropriate space of bundles. The minimal Hecke operators indeed 
satisfy the oper equation with a classical stress tensor built from the Hitchin Hamiltonians.

We should recall how the identification of ${\cal M}(z)$ Ward identities with Hecke operators is derived. 
The crucial observation is that ${\cal M}(z)$ can be obtained as an average of spectral flow operators. 

Indeed, consider the simplest $Z^{\pm\beta\gamma}(z) \to z^{\pm \frac12}Z^{\pm\beta\gamma}(z)$ spectral flow. It maps the vacuum to 
a highest weight Ramond vertex operator ${\cal S}_0$ which is annihilated by all the $Z^{- \beta \gamma}_0$ zeromodes. 
A slightly more general spectral flow operation would give an highest weight Ramond vertex operator ${\cal S}_\mu$ which is annihilated by all the $Z^{- \beta \gamma}_0 + \mu Z^{+ \beta \gamma}_0 $ zeromodes. It is easy to show that we can write ${\cal S}_\mu$ as an exponentiated $J^{++}$ rotation of ${\cal S}_0$:
\begin{equation}
{\cal S}_\mu = e^{\frac12 \mu \det_{\beta \gamma} Z^{+\beta \gamma}_0} \circ {\cal S}_0
\end{equation}
Averaging over $\mu$ gives:
\begin{equation}
\oint d\mu {\cal S}_\mu = \delta(\det_{\beta \gamma} Z^{+\beta \gamma}_0) \circ {\cal S}_0
\end{equation}
and thus we recognize the integral transform as producing ${\cal M}(z)$. The same construction applied to the other 
$PSL(2)$ groups gives different integral transform constructions for the same vertex operator ${\cal M}(z)$.  

In the non-chiral setup the contour integral is replaced by an integral over the $\mu$ plane. We can denote the special Ramond vertex operator as ${\mathbb M}(z)$ 
and write
\begin{equation}
{\mathbb M}(z) = \int d\mu d \bar \mu\, {\mathbb S}_{\mu} 
\end{equation}

A correlation function including some even number of ${\cal M}(z)$ insertions and some other vertex operators such as the ${\cal V}(x_i,x'_i,x''_i;z_i)$ can be written as an integral over correlation functions with some ${\cal S}_\mu$ insertions. The latter insertions can be then eliminates pairwise by spectral flows, 
resulting in some integral operators acting on the correlation function with no ${\cal M}(z)$ insertions, which are the Hecke operators. 

We can also introduce another convenient class of Ramond vertex operators, which represent the action of Hecke operators at a pre-existent puncture. 
Consider an highest weight Ramond vertex operator ${\cal W}(x,x',x'';z)$ which is annihilated by all zeromodes of the form $x_\alpha x'_\beta Z^{\alpha \beta \gamma}_0$, 
$x_\alpha x''_\gamma Z^{\alpha \beta \gamma}_0$ or $ x'_\beta x''_\gamma Z^{\alpha \beta \gamma}_0$. This is a four-dimensional Lagrangian subspace, 
so it is a sensible definition. We can visualize (and normalize) these operators as an $SL(2)^3$ rotation of a vertex operator ${\cal W}(0,0,0;z)$ annihilated by $Z^{---}_0$, $Z^{+--}_0$, $Z^{-+-}_0$ and $Z^{--+}_0$. 

These operators have the same properties as the ${\cal V}(x,x',x'';z)$ under the action of the $\mathfrak{sl}(2)^3$ Kac-Moody currents. 
Indeed, the two are related by a minimal spectral flow operation involving either of the three $SL(2)$'s. For example, start from the ${\cal W}(0,0,0;z)$  vertex operator annihilated by $Z^{---}_0$, $Z^{+--}_0$, $Z^{-+-}_0$ and $Z^{--+}_0$. This is also annihilated by all of the $Z^{\alpha \beta \gamma}_1$ because of the highest weight condition. A spectral flow under the first $SL(2)$ maps it to a state annihilated by 
$Z^{---}_{\frac12}$, $Z^{+--}_{-\frac12}$, $Z^{-+-}_{\frac12}$ and $Z^{--+}_{\frac12}$ as well as all of the $Z^{+ \beta \gamma}_{\frac12}$. This is 
the same as ${\cal V}(\infty,0,0)$. More generally, the same spectral flow maps ${\cal W}(0,x',x'')$ to ${\cal V}(\infty,x',x'')$, so the two must have 
the same properties under the last two $SL(2)$'s. Etcetera. 

Analogous statements apply to the non-chiral versions ${\mathbb W}(x,x',x'';z)$ of the ${\cal W}(x,x',x'';z)$.
We thus have a collection of correlation functions which can give, in the non-chiral setup, half-densities on ${\cal M}_0(\mathrm{PSL}(2),\mathbb{CP}^1_n)$. 
A sphere correlation function of ${\mathbb V}(x_i,x'_i,x''_i;z_i)$ give the basic intertwining kernels $\mathbb{K}_3$. 
Replacing some ${\mathbb V}(x_i,x'_i,x''_i;z_i)$ by ${\cal W}(x_i,x_i',x_i'';z_i)$ give wavefunctions which can be obtained from $\mathbb{K}_3$
by special Hecke modifications at $z_i$. Adding ${\cal M}(z)$ give wavefunctions which can be obtained from $\mathbb{K}_3$ by minimal Hecke modifications 
at $z$, etc. 

\subsection{Some explicit formulae}
It is easy to reproduce in this language the complex Hecke operators from \cite{Etingof:2021eeu}. We take $m+2$ operators ${\mathbb V}$ on the sphere and 
for convenience we place ${\mathbb V}(x_{m+1},x_{m+1}',x_{m+1}'';z_{m+1})$ at infinity both in the $z$ plane and in the $x$, $x'$ and $x''$ variables. 
When computing the kernel, that means $A_{i,m+1} = A_{m+1,i}=1$. 

It may also be convenient to set $z_0=x_0=x'_0=x''_0=0$ and $z_1=x_1=x'_1=x''_1=1$ to completely fix the global $PSL(2)$ symmetries. 
For example, the kernel for four points becomes 
\begin{align}
&\mathbb{K}_3(x,x',x'';z)= \cr
&\frac{|z|^2 |1-z|^2}{|(z (z - 1) + (1 - x) (1 - x') (1 - x'') z + x x' x'' (1 - z))^2 + 4xx'x''(x-1)(x'-1)(x''-1)z (z - 1)|}\cr&
\end{align}
The four Gaudin Hamiltonians collapse to the Lame' operator
\begin{equation}
L = \partial_x x(x-1)(x-z) \partial_x + x
\end{equation}
and we can easily verify that 
\begin{equation}
L \circ \mathbb{K}_3(x,x',x'';z)=L' \circ \mathbb{K}_3(x,x',x'';z)=L'' \circ \mathbb{K}_3(x,x',x'';z)
\end{equation}
and analogous anti-holomorphic relations. 

Now consider replacing ${\mathbb V}(\infty, \infty, \infty;\infty)$ with ${\mathbb W}(\infty, \infty, \infty;\infty)$ and ${\mathbb V}(0, 0, 0;0)$ with ${\mathbb W}(0, 0, 0;0)$.
The resulting correlation function can be simplified by applying a spectral flow Ward identity for the transformation $Z^{\pm \beta \gamma} \to z^{\pm \frac12}Z^{\pm \beta \gamma}$, which converts ${\mathbb W}(0, 0, 0;0)$ to ${\mathbb V}(\infty, 0, 0;0)$ and ${\mathbb W}(\infty, \infty, \infty;\infty)$ to ${\mathbb V}(0, \infty, \infty;\infty)$. At the location of other operators, the field redefinition maps ${\mathbb V}(x_i, x'_i,x''_i;z_i) \to {\mathbb V}(x_i/z_i, x'_i,x''_i;z_i)\prod_i |z_i|^{-1}$.

We can rotate back $x_0$ and $x_{\infty}$ to the standard values by an inversion $x_i \to \frac{1}{x_i}$ which maps the other vertex operators to 
${\mathbb V}(z_i/x_i, x'_i,x''_i;z_i)\prod_i  |z_i| |x_i|^{-2}$. The transformation $x_i \to z_i/x_i$ accompanied by a rescaling by $|z_i| |x_i|^{-2}$ 
is called $S_0$ in reference \cite{Etingof:2021eeu}: a simultaneous special Hecke modification at $0$ and $\infty$. We easily verify the somewhat non-trivial relations 
\begin{equation}
S_0 \circ \mathbb{K}_3(x,x',x'';z)=S_0' \circ \mathbb{K}_3(x,x',x'';z)=S_0'' \circ \mathbb{K}_3(x,x',x'';z)
\end{equation}
We can define and test in a similar manner the $S_i$ transformations acting at $i$ and $\infty$ by special Hecke transformations. 

Next, we can both replace ${\mathbb V}(\infty, \infty, \infty;\infty)$ with ${\mathbb W}(\infty, \infty, \infty;\infty)$ and insert ${\mathbb M}(z)$ in the correlation function. 
This will describe the action of an Hecke operator at $z$.  It is now convenient to allow generic values of $z_0, x_0$, etc. and $z_1, x_1,$ etc. 

We simplify the resulting correlator in a few steps. First, we express ${\mathbb M}(z)$ as an average of the spectral flow operators ${\mathbb S}(z)_{\mu}$.
Next, we apply a translation in the first $PSL(2)$ factor to map that insertion to ${\mathbb S}(z)_{0}$. This changes the argument of the 
other insertions to $x_i - \mu$. A spectral flow transformation then eliminates ${\mathbb S}(z)_{0}$
and maps ${\mathbb W}(\infty, \infty, \infty;\infty)$ to ${\mathbb V}(0, \infty, \infty;\infty)$ while transforming the argument of the ${\mathbb V}$ insertions to 
$\frac{x_i - \mu}{z_i-z}$. Finally, an inversion brings the vertex operator at infinity to the canonical ${\mathbb V}(\infty, \infty, \infty;\infty)$
and the other insertions get argument $-\frac{z_i-z}{x_i - \mu}$. 

The final result is an average over the $\mu$ plane of the ${\mathbb V}$ correlation function evaluated at $x_i \to -\frac{z_i-z}{x_i - \mu}$. 
This is the Hecke operator at position $z$ from Proposition $3.9$ in reference \cite{Etingof:2021eeu}. Our chiral algebra analysis 
thus verifies that ${\mathbb K}_3$ intertwines the action of the Hecke operators on the three sets of variables. 

A direct check of the relationship in a simple example such as the four-punctured sphere is cumbersome but not impossible. 
The action of the Hecke operator on the kernel ${\mathbb K}_3$ involves an integral over the $\mu$ plane of the inverse modulus of 
a degree four polynomial $P_4(\mu)$ in $\mu$. The answer is a bilinear of the periods of the differential $\frac{d\mu}{\nu}$ 
on the elliptic curve 
\begin{equation}
\nu^2 = P_4(\mu)
\end{equation}
and can be expressed in terms of its $g_2$ and $g_3$ invariants. It is easy to check that these two quantities are invariant under $x_i \leftrightarrow x'_i$, 
which essentially implies the desired intertwining relation. 

This explicit proof strategy can in principle be extended to any number of points on the sphere. It is also sufficiently algebraic that
it could perhaps be extended to other fields besides $\mathbb{C}$ and $\mathbb{R}$.

\section{The $SO(2N+1) \subset SO(2N+2)$ dual kernel} \label{sec:emb}
In this section we consider the theory of symplectic bosons $Z^{\alpha a}$, with $\alpha = 1,\cdots 2N$ and $a=1, \cdots, 2N+2$ and symplectic form 
\begin{equation}
\omega^{\alpha \beta} \delta^{ab}
\end{equation}
This chiral algebra has an $\mathrm{sp}(2N) \times \mathrm{so}(2N+2)$ Kac-Moody subalgebra at critical level. The case $N=1$ corresponds to the trifundamental 
symplectic bosons from the previous section. 

The centers of the two Kac-Moody subalgebras are not independent. We expect them to satisfy 
a relationship which is Langlands dual to the natural embedding of $SO(2N+1)$ opers into $SO(2N+2)$ opers. For $N=1$, this is precisely the relation that the 
three quadratic Casimirs coincide. \footnote{Our expectation is motivated by one of the S-duality statements from \cite{Gaiotto:2008ak}. The VOA at hand 
is associated to a set of bi-fundamental 3d (half)hypermultiplets, defining an interface between $SP(2N)$ and $SO(2N+2)$ gauge theories. That interface is S-dual to an interface between $SO(2N+1)$ and $SO(2N+2)$ gauge theories, such that the $SO(2N+1)$ gauge group at the embeds block-diagonally in $SO(2N+2)$ at the interface.  }

In order to define useful sphere correlation functions we can employ pure vectors of the form $x_\alpha v_a$, where $x_\alpha$ is generic but $v_a$ is a
null vector in $\mathbb{C}^{2N+2}$. Again, for $N=1$ this definition agrees with what we did for the trifundamental case. We can thus produce vertex operators 
${\cal V}(x,v;z)$ labelled by a point $[x]$ in ${\mathbb CP}^{2N-1}$ and a point $[v]$ in the projective null cone in ${\mathbb CP}^{2N+1}$.

These vertex operators are Kac-Moody primaries and thus the insertion of central elements in a sphere correlation function will be expressed by the action of 
$\mathrm{sp}(2N) \times \mathrm{so}(2N+2)$ Gaudin Hamiltonians, built from the action on ${\mathbb CP}^{2N-1}$ and on the projective null cone respectively. 

The genus $0$ correlation function defines the chiral kernel 
\begin{equation}
K(x_*,v_*;z_*) \equiv \left\langle \prod_{i=1}^n {\cal V}(x_i,v_i;z_i) \right\rangle_{\mathbb{CP}^1} = \frac{1}{\sqrt{\det A}}
\end{equation}
computed from the matrix 
\begin{equation}
A^{ij} = \frac{(x_i,x_j)(v_i,v_j)}{z_i-z_j} \qquad \qquad i \neq j
\end{equation}
and $A^{ii}=0$.
We thus predict that this kernel will satisfy differential constraints expressing the $SO(2N+1) \subset SO(2N+2)$ embedding of the Gaudin Hamiltonians
for $\mathrm{sp}(2N) \times \mathrm{so}(2N+2)$.

\section{Non-chiral BRST reduction} \label{sec:brst}
We will now broaden the class of chiral algebras we consider. Take a collection of symplectic bosons in a representation $V$ of a reductive group $G$ 
such that the $\mathfrak{g}$ Kac-Moody currents have a level which is twice the critical value. We can then consider a chiral 2d $G$ gauge theory 
coupled to the symplectic bosons. In practice, we add $bc$ ghosts valued in $\mathfrak{g}$ and do a BRST reduction using a canonical 
BRST charge of the schematic form $bcc + c J$, such that $\{Q,b\} = J$. 

The result will be a new (possibly derived) chiral algebra which we can denote as $\mathrm{Sb}^{V/G}$. It is a chiral algebra version of a complex symplectic quotient 
of $V$ by $G$. We can obtain modules for $\mathrm{Sb}^{V/G}$ by taking a BRST reduction of the corresponding modules for $\mathrm{Sb}^{V}$. 
Modules related by a spectral flow associated to $G$ have the same image under the BRST reduction. If some other reductive group $G'$ 
has an action on $V$ commuting with the $G$ action, then $\mathrm{Sb}^{V/G}$ has a $G'$  Kac-Moody symmetry with the same level as $\mathrm{Sb}^{V}$.

The vacuum module of $\mathrm{Sb}^{V}$ descends, by definition, to the vacuum module of $\mathrm{Sb}^{V/G}$. Highest weight Ramond modules 
also have a nice image, as they are also highest weight under $G$. That means that a highest weight Ramond vertex operator is BRST-closed as long as it is invariant under the action of the $G$ Kac-Moody zeromodes. Concretely, we have a module $M$ for the Weyl algebra associated to $V$ and we are looking for vectors fixed by the corresponding $G$ action on $M$. 

Correlation functions of $\mathrm{Sb}^{V/G}$ can be computed schematically as follows. Correlation functions of $\mathrm{Sb}^{V}$ can be coupled to $G$-bundles 
and transform as (possibly singular) sections of the canonical bundle on the space of bundles $\mathrm{Bun}_G$. The singularities occur whenever the symplectic bosons have zeromodes. We can define correlation functions of $\mathrm{Sb}^{V/G}$ by integrating correlation functions of $\mathrm{Sb}^{V}$
on middle-dimensional cycles in $\mathrm{Bun}_G$. The choice of integration cycles introduces further multi-valuedness in the correlation functions. 

A crucial question for our applications is if a non-chiral version of this construction exists, as described in \cite{Gaiotto:2021tsq}, and may provide 
a true 2d CFT $T_\mathrm{Sb}^{V/G}$. The rationale for the construction is that a path integral over unitary $G$ connections coupled to matter which is Kac-Moody invariant can be simplified to a finite-dimensional integral over $\mathrm{Bun}_G$, up to an overall factor. This is completely analogous to a fact familiar from bosonic string theory: a path integral over 2d metrics coupled to Weyl-invariant matter can be reduced to an integral over the moduli space of complex structures. 

We expect a 2d gauge theory with unitary gauge group coupled to the non-chiral version $T_\mathrm{Sb}^{V}$ of symplectic bosons to give a well-defined 2d CFT 
$T_\mathrm{Sb}^{V/G}$. The correlation functions of $T_\mathrm{Sb}^{V}$ give densities on $\mathrm{Bun}_G$ and the correlation functions of $T_\mathrm{Sb}^{V/G}$ will be actual integrals of these densities over $\mathrm{Bun}_G$. An important challenge is that the correlation functions of $T_\mathrm{Sb}^{V}$ diverge on a 
``theta divisor'' where zeromodes appear. The correlation functions of $T_\mathrm{Sb}^{V/G}$ can only be defined if that divergence is integrable. The classical equations of motion for the 2d gauge fields set to zero the $G$ Kac-Moody currents. We thus suspect that 
divergences of $T_\mathrm{Sb}^{V/G}$ correlation functions will occur when the symplectic boson zeromodes can be tuned not to induce classical vevs of the $G$ Kac-Moody currents. 

If we consider sphere correlation functions, the role of $\Bun_G$ is played by whatever parameters control the action of the positive modes of the $G$ currents on the vertex operators we select for the $T_\mathrm{Sb}^{V}$ correlation function, modulo global $G$ rotations. If we can average the vertex operators to make them invariant under the action of these positive modes, the resulting vertex operator will be BRST invariant. Up to some subtleties involving gauge-fixing the global $G$ symmetry, we can use these integrated vertex operators to obtain a correlation function in $T_\mathrm{Sb}^{V/G}$. 
 
\subsection{The $V_{\mathfrak{sl}(2),4}$ chiral algebra and $T_{\mathfrak{sl}(2),4}$ 2d CFT}
Our main example of interest is associated to kernels $K_4$ and non-chiral versions which intertwine the action of $PSL(2)$ Hitchin Hamiltonians and Hecke operators for four sets of variables. The corresponding chiral algebra $V_{\mathfrak{sl}(2),4}$ is obtained as the BRST reduction of $V_{\mathfrak{sl}(2),3} \times V_{\mathfrak{sl}(2),3}$ by a diagonal $SL(2)$ subgroup, i.e. it is the same as $\mathrm{Sb}^{\mathbb{C}^{18}/SL(2)}$. 

The chiral algebra has an $\mathrm{Spin}(8)$ global symmetry, with the underlying $SL(2)^4$ subgroup appearing as a block-diagonal $\mathrm{Spin}(4)^2 \subset \mathrm{Spin}(8)$. This description of the algebra hides an important fact: the four $SL(2)$ subgroups appear on the same footing, as 
 $V_{\mathfrak{sl}(2),4}$ is invariant under triality automorphisms of $\mathrm{Spin}(8)$. This symmetry only appears after the BRST reduction and is 
 broken by the $\mathrm{Sb}^{\mathbb{C}^{18}/SL(2)}$ description. 
 
The chiral algebra $V_{\mathfrak{sl}(2),4}$ is actually generated by the $\mathrm{spin(8)}$ Kac-Moody currents, but it is a quotient of the full Kac-Moody algebra. 

Schematically, we can produce the kernel $K_4(x,x',x'',x''')$ as an integral over $y$ of the product
\begin{equation}
K_3(x,x',y)K_3(y,x'',x''').
\end{equation} The only subtlety is that we should 
gauge-fix three of the $y_i$ variables, say to $0$,$1$,$\infty$, and integrate over the remaining ones. As long as the integral is sensible, 
integration by parts makes manifest that $K_4(x,x',x'',x''')$ intertwines the action of Gaudin Hamiltonians and Hecke operators for 
all four sets of variables. 

A property which is {\it not} manifest but appears to be true experimentally is that $K_4(x,x',x'',x''')$ is symmetric under permutations of the four sets of variables. 
We expect this property to follow from a symmetry property of the corresponding averaged vertex operators
\begin{equation}
 {\cal V}_4(x,x',x'',x''';z) \equiv \oint dy\,  {\cal V}(x,x',y;z) \otimes {\cal V}(x'',x''',y;z)
\end{equation}
Formally, we can write 
\begin{equation}
 {\cal V}_4(0,0,0,0;z) \equiv \oint dy d\lambda d\tilde \lambda \,  e^{\lambda (Z^{---} + y Z^{--+})+\tilde \lambda (\tilde Z^{---} + y \tilde Z^{--+})}
\end{equation}
giving 
\begin{equation}
 {\cal V}_4(0,0,0,0;z) \equiv \oint  d\lambda \,  e^{\lambda (Z^{---}\tilde Z^{--+} -\tilde Z^{---}Z^{--+} )}=\oint  d\lambda \,  e^{\lambda \epsilon_{\alpha \beta} Z^{--\alpha}\tilde Z^{--\beta} }
\end{equation}
We can write this in terms of the Kac-Moody currents 
\begin{equation}
J^{\alpha \beta \gamma \delta} = \epsilon_{\alpha' \beta'}Z^{\alpha \beta\alpha'}\tilde Z^{\gamma \delta\beta'}
\end{equation}
as
\begin{equation}
 {\cal V}_4(x,x',x'',x''';z) \equiv \oint  d\lambda \,  e^{\lambda x_\alpha x'_\beta x''_\gamma x'''_\delta J^{\alpha \beta \gamma \delta}}
\end{equation}
This expression only involves generators of $V_{\mathfrak{sl}(2),4}$ and is manifestly symmetric under permutation of the four $SL(2)$ subgroups.

Incidentally, this expression and its derivation can be generalized to define the relevant vertex operators in more complicated $V_{\mathfrak{sl}(2),n}$
examples. These algebras have a generator $W^{\alpha_1, \cdots \alpha_n}$ built in a manner analogous to $J^{\alpha \beta \gamma \delta}$
and we could write
\begin{equation}
 {\cal V}_4(x^1,\cdots x^n;z) \equiv \oint  d\lambda \,  e^{\lambda x^1_{\alpha_1} \cdots x^n_{\alpha_n}  W^{\alpha_1, \cdots \alpha_n}}
\end{equation}

It would be interesting to derive analogous vertex operators in more general $V_{\mathfrak{g},n}$.

\subsection*{Acknowledgements}
This research was supported in part by a grant from the Krembil Foundation. D.G. is supported by the NSERC Discovery Grant program and by the Perimeter Institute for Theoretical Physics. Research at Perimeter Institute is supported in part by the Government of Canada through the Department of Innovation, Science and Economic Development Canada and by the Province of Ontario through the Ministry of Colleges and Universities. 

\bibliography{Langlands}{}

\providecommand{\href}[2]{#2}\begingroup\raggedright\begin{thebibliography}{10}

\bibitem{Teschner:2017djr}
J.~Teschner, \emph{{Quantisation conditions of the quantum Hitchin system and
  the real geometric Langlands correspondence}},
  \href{https://arxiv.org/abs/1707.07873}{{\ttfamily 1707.07873}}.

\bibitem{Etingof:2019pni}
P.~Etingof, E.~Frenkel and D.~Kazhdan, \emph{{An analytic version of the
  Langlands correspondence for complex curves}},
  \href{https://arxiv.org/abs/1908.09677}{{\ttfamily 1908.09677}}.

\bibitem{Etingof:2021eub}
P.~Etingof, E.~Frenkel and D.~Kazhdan, \emph{{Hecke operators and analytic
  Langlands correspondence for curves over local fields}},
  \href{https://arxiv.org/abs/2103.01509}{{\ttfamily 2103.01509}}.

\bibitem{Etingof:2021eeu}
P.~Etingof, E.~Frenkel and D.~Kazhdan, \emph{{Analytic Langlands correspondence
  for PGL(2) on P\textasciicircum{}1 with parabolic structures over local
  fields}},  \href{https://arxiv.org/abs/2106.05243}{{\ttfamily 2106.05243}}.

\bibitem{Gaiotto:2021tsq}
D.~Gaiotto and E.~Witten, \emph{{Gauge Theory and the Analytic Form of the
  Geometric Langlands Program}},
  \href{https://arxiv.org/abs/2107.01732}{{\ttfamily 2107.01732}}.

\bibitem{Gaiotto:2016wcv}
D.~Gaiotto, \emph{{Twisted compactifications of 3d $ \mathcal{N} $ = 4 theories
  and conformal blocks}},
  \href{https://doi.org/10.1007/JHEP02(2019)061}{\emph{JHEP} {\bfseries 02}
  (2019) 061} [\href{https://arxiv.org/abs/1611.01528}{{\ttfamily
  1611.01528}}].

\bibitem{2007math......2206K}
M.~{Kontsevich}, \emph{{Notes on motives in finite characteristic}},
  {\emph{arXiv Mathematics e-prints} (2007) math/0702206}
  [\href{https://arxiv.org/abs/math/0702206}{{\ttfamily math/0702206}}].

\bibitem{2018arXiv181101577A}
T.~{Arakawa}, \emph{{Chiral algebras of class $\mathcal{S}$ and Moore-Tachikawa
  symplectic varieties}}, {\emph{arXiv e-prints} (2018) arXiv:1811.01577}
  [\href{https://arxiv.org/abs/1811.01577}{{\ttfamily 1811.01577}}].

\bibitem{Costello:2018fnz}
K.~Costello and D.~Gaiotto, \emph{{Vertex Operator Algebras and 3d $
  \mathcal{N} $ = 4 gauge theories}},
  \href{https://doi.org/10.1007/JHEP05(2019)018}{\emph{JHEP} {\bfseries 05}
  (2019) 018} [\href{https://arxiv.org/abs/1804.06460}{{\ttfamily
  1804.06460}}].

\bibitem{Gaiotto:2008ak}
D.~Gaiotto and E.~Witten, \emph{{S-Duality of Boundary Conditions In N=4 Super
  Yang-Mills Theory}},
  \href{https://doi.org/10.4310/ATMP.2009.v13.n3.a5}{\emph{Adv. Theor. Math.
  Phys.} {\bfseries 13} (2009) 721}
  [\href{https://arxiv.org/abs/0807.3720}{{\ttfamily 0807.3720}}].

\end{thebibliography}\endgroup
\bibliographystyle{JHEP}
\end{document}